# Monte Carlo simulation of the secondary electron yield of silicon rich silicon nitride

A.M.M.G. Theulings[a], S.X. Tao[b], C.W. Hagen[a] and H. van der Graaf [a,c,1,2]

[a] *Faculty of Applied Sciences, Department of Imaging Physics, Delft University of Technology, Lorentzweg 1. 2628 CJ, Delft, The Netherlands*
[b] *now at: Materials Simulation & Modelling, Department of Applied Physics, Eindhoven University of Technology, 5600 MB Eindhoven, The Netherlands*
[c] *Nikhef, Science Park 106, Amsterdam, The Netherlands*
[1] *Retired*
[2] *Corresponding author: E-mail* `vdgraaf@nikhef.nl`

ABSTRACT: The effect of doping in $Si_3N_4$ membranes on the secondary electron yield is investigated using Monte Carlo simulations of the electron-matter interactions. The effect of the concentration and the distribution of the doping in silicon rich silicon nitride membranes is studied by using the energy loss function as obtained from ab initio density functional theory calculations in the electron scattering models of the Monte Carlo simulations. An increasing doping concentration leads to a decreasing maximum secondary electron yield. The distribution of the doped silicon atoms can be optimised in order to minimize the decrease in yield.

# 1. Introduction

Photomultiplier tubes are in use since their invention in the 1930s due to their efficiency, time resolution and low noise characteristics. The one application where they cannot be used is in a (dynamic) magnetic field. The **Ti**med **P**hoton **C**ounter (TiPC) proposed in [1] [2] [3] is a single photon detector with high spatial and time resolution that does have the ability to work in dynamic magnetic fields. The TiPC consists of a stack of thin transmission dynodes (tynodes) on top of a pixel chip, the whole is capped by a photocathode. When the photocathode collects a photon, the emitted photoelectron is accelerated towards the first tynode. This tynode is a thin membrane such that a high energy incoming electron from the top results in multiple low energy electrons emitted at the bottom. Now the low energy emitted electrons are accelerated to the second tynode and the multiplication process repeats. After the last tynode, the electrons hit a pixel input pad on the pixel chip and the signal is detected. The pixel input pads of a Timepix chip can detect signals starting from 1000 electrons [8]. A feasible amount of tynodes to stack on top of the pixel chip would be five tynodes. In order to have a signal of 1000 electrons, the multiplication per tynode should be at least 4.

Especially for the last tynode, where the absolute number of electrons emitted is the largest, charge up effects may become important. The charging of a tynode will cause the secondary electron yield (SEY) to decrease, an unwanted effect. A solution to the charging problem would be adding a dopant to the material to make the material slightly conductive. An example of this is boron doping in diamond [4]. However, diamond membranes that are thin enough to be used as tynodes are hard to fabricate. A suitable material to make thin membranes of, and to which dopants can be added, is silicon nitride [5] [6].

The effect of silicon doping in silicon nitride membranes on the SEY can be investigated with Monte Carlo simulations of the electron-matter interactions. Because the doping counteracts charge up effects, the Monte Carlo simulator does not take charging into account.

Five different forms of (silicon rich) silicon nitride (SRSN) were investigated: $Si_3N_4$, $Si_{13}N_{15}$, homogeneous $Si_7N_7$, clustered $Si_7N_7$ and amorphous $Si_7N_7$. The difference between the three forms of $Si_7N_7$ that were investigated is the distribution of the extra silicon atoms. Here 'extra' is defined as the silicon atoms that need to be 'added' to $Si_3N_4$ to obtain heavily doped silicon nitride that it becomes $Si_7N_7$. In homogeneous $Si_7N_7$, the extra silicon atoms are distributed homogeneously over the material and in clustered $Si_7N_7$, the extra silicon atoms are placed in clusters that are distributed homogeneously over the material. It is expected that lattice defects are introduced when extra silicon atoms are introduced in pure silicon nitride, to obtain SRSN. These lattice defects are effectively dangling silicon bonds of the extra silicon atoms. This work focuses on the effect of the different forms of SRSN containing various forms of dangling bonds and their effect on the secondary electron yield.

# 2. Simulation models and material parameters

The Monte Carlo simulation package used to obtain the results is based on the electron-matter interaction code developed by Kieft and Bosch [7] and the modifications made by Verduin [8].

The models used describe the interactions between the energetic electrons and the ambient material. Three types of interactions are distinguished: elastic interactions, where the electron only changes direction, inelastic interactions, where the electron also loses energy to the material and the boundary crossing between vacuum and matter. The differences with the original models as developed by Kieft and Bosch are:

1. Above 200 eV the elastic interactions are modelled using Mott cross sections. Kieft and Bosch did not take solid state effects into account. In this work they are taken into account. Below 200 eV the elastic interaction is modelled using acoustic phonon scattering. In the original models, only longitudinal acoustic phonons are taken into account. In this work the two transversal modes are also taken into account. Both these improvements result in more accurate cross sections.
2. The inelastic scattering mean free path is calculated using the dielectric function theory. Some errors in the original models were corrected and a phenomenological factor of 1.5 was removed.
3. In the models that govern the boundary crossing between two materials, there was an error in the calculation of the potential step for insulators and semiconductors in the original models, which was corrected in this work.

More details about the models used in the simulation package can be found in [9].
In order to simulate a material, the material parameters need to be known. Most material parameters were very straightforward to find and are listed in Table 1, including the mass density, the electron affinity, the atomic mass of silicon and nitrogen and the sound velocity. However, some material parameters were more difficult to obtain and deserve a few words of explanation. When using the dielectric function theory, the energy loss function of the material is needed. For the different forms of SRSN investigated here the energy loss function was not known. However, it is possible to calculate the energy loss function with the use of *ab initio* density functional theory calculations [10]. Although not reported in [10], the authors kindly provided the dielectric function of amorphous $Si_7N_7$ also. Approximate values of the band gap for the different forms of SRSN were taken from the density of states as calculated in [10] as well.
The last parameter that was difficult to obtain was the acoustic deformation potential, which is necessary to calculate the acoustic phonon scattering. For materials for which a value could be found, the acoustic deformation potential typically lies between 2 and 15 eV [11].
The definition of the acoustic deformation $\epsilon_{ac}$ potential is
$$\delta E_c = \epsilon_{ac} \frac{\delta V}{V},$$
where $\delta E_c$ is the absolute energy shift of the conduction band minimum for a small uniform expansion $\delta V$ of the crystal. When looking at the density of states for the different doping levels in [10] it is concluded that the effect of the extra silicon atoms on the band structure is to introduce defect states. However, the general shape of the band structure is not affected. Since this is the case, we do not expect $\delta E_c$ to change significantly when extra silicon atoms are introduced. Hence it is assumed that the same value for the acoustic deformation potential can be used for the different forms of SRSN. Tekippe [12] investigated the effect of different kinds of dopant in silicon on the acoustic deformation potential and they found that the acoustic deformation potential did indeed not depend on the kind of dopant. It is therefore concluded that a single value for the acoustic deformation potential can be chosen for the different forms of SRSN and the results can be compared qualitatively. Hence the value of 12.0 eV is chosen in this work.

**Table 1.** Material properties used in the simulations

|  | $Si_3N_4$ | $Si_{13}N_{15}$ | $Si_7N_7$ | | |
|---|---|---|---|---|---|
|  |  |  | homogeneous | clusters | amorphous |
| Mass density | 3.27 g/cm$^3$ [13] | 3.27 g/cm$^3$ [13] | 3.27 g/cm$^3$ [13] | 3.27 g/cm$^3$ [13] | 3.27 g/cm$^3$ [13] |
| Fermi energy | 7.56 eV [10] | 7.84 eV [10] | 9.8 eV [10] | 9.98 eV [10] | 9.54 eV [14] |
| Band gap | 4.7 eV [10] | 4.4 eV [10] | 3.9 eV [10] | 3.6 eV [10] | 3.5 eV [14] |
| Electron affinity | 1.5 eV [7] | 1.5 eV [7] | 1.5 eV [7] | 1.5 eV [7] | 1.5 eV [7] |
| Lattice constant | 5.26 Å [10] | 6.81 Å [10] | 10.5 Å [10] | 10.53 Å [10] | 10.66 Å [14] |
| Atomic mass Si | 28.0855 g/mole [15] | 28.0855 g/mole [15] | 28.0855 g/mole [15] | 28.0855 g/mole [15] | 28.0855 g/mole [15] |
| Atomic mass N | 14.007 g/mole [15] | 14.007 g/mole [15] | 14.007 g/mole [15] | 14.007 g/mole [15] | 14.007 g/mole [15] |
| Sound velocity | 7833 m/s [13] | 7833 m/s [13] | 7833 m/s [13] | 7833 m/s [13] | 7833 m/s [13] |
| $\epsilon_{ac}$ | 12.0 eV | 12.0 eV | 12.0 eV | 12.0 eV | 12.0 eV |

## 3. Simulation results and discussion

**Doping concentration.** In SRSN, there are extra silicon atoms present compared to pure silicon nitride. These extra silicon atoms will introduce lattice defects in the form of dangling bonds. Electrons travelling in the material can scatter at these defects and lose energy. It is expected to see this in the energy loss function in the form of additional energy loss peaks. In Figure 1 the calculated energy loss functions for pure silicon nitride ($Si_3N_4$), $Si_{13}N_{15}$ and $Si_7N_7$ are shown. The extra silicon atoms are distributed homogeneously over the material for both $Si_{13}N_{15}$ and $Si_7N_7$. Additional energy loss peaks are seen in the energy loss function for SRSN compared to pure silicon nitride, as expected. The energy loss function for $Si_{13}N_{15}$ shows extra peaks around 1.5 eV and 3.0 eV. Introducing even more silicon atoms results in loss peaks around 0.8 eV, 1.7 eV and 5 eV for $Si_7N_7$. Note that below 10 eV, the energy loss function increases overall for increasing doping level. Due to the overall increase and the extra energy loss peaks below 10 eV, the secondary electron yield is expected to decrease with increasing doping level. Intuitively, this can be understood by considering that the extra silicon atoms in the material introduce dangling bonds. These dangling bonds form potential inelastic scattering sites for (secondary) electrons travelling in the material. Each dangling bond perturbs the potential energy landscape from the 'equilibrium' potential energy landscape if no silicon doping would be present. Electrons travelling in the material will have a bigger probability of scattering inelastically due to this perturbation, or could even be absorbed into the material at such a perturbation. Indeed, in Figure 2, the simulated secondary electron yield decreases with increasing doping level. The maximum secondary electron yield of $Si_{13}N_{15}$ is 3.3 at 350 eV, a decrease of 34% compared to 5.0 at 450 eV for $Si_3N_4$. When increasing the doping level even more to get $Si_7N_7$, the maximum secondary electron yield decreases further to 2.7 at 350 eV, a total decrease of 46%. Although the secondary electron yield decreases with increasing doping level, it might still be beneficial to have at least some silicon doping in samples used for experiments. During experiments, special precautions need to be taken to prevent the samples from charging. The silicon doping in SRSN makes the material slightly more conductive than pure silicon nitride and this helps in preventing charge up effects [16].

**Dopant distribution.** So far only SRSN was discussed, where the extra silicon atoms are distributed homogeneously over the material with the same crystal structure as pure silicon nitride. In this case, every additional silicon atom introduces one dangling bond, as it replaces one nitrogen atom in the crystal structure. Other distributions of the additional silicon atoms are also possible. One option would be that the extra silicon atoms group together in clusters. In these clusters, at least some of the dangling bonds can be removed i.e., two silicon atoms both with a dangling bond can bond to each other, thus removing two dangling bonds. As a result, the energy loss peaks due to the extra silicon atoms are expected to decrease for such a dopant distribution. Indeed, when comparing the energy loss function of $Si_7N_7$ with silicon clusters to that of homogeneous $Si_7N_7$ in Figure 3, the energy loss peaks below 10 eV are seen to decrease. The energy loss function of $Si_7N_7$ with silicon clusters has energy loss peaks at 0.6 eV, 2.5 eV, 3.4 eV. The energy loss peaks are such that in between 2 eV and 10 eV, the energy loss functions for homogeneous and clustered $Si_7N_7$ are comparable and below 2 eV, the energy loss function for clustered $Si_7N_7$ is lower than that for homogeneous $Si_7N_7$. A third possibility is amorphous $Si_7N_7$. In this case, all atoms are allowed to relax and possibly even more of the dangling silicon bonds can be removed. The energy loss function for amorphous $Si_7N_7$ in Figure 3 indeed shows fewer low energy loss peaks, which seems to suggest that more silicon dangling bonds are removed. The energy loss function of amorphous $Si_7N_7$ has only one real energy loss peak left at 2.5 eV. For both clustered and amorphous $Si_7N_7$ the bulk energy loss peak is shifted from 24 eV to 20 eV and becomes narrower. In fact, both peaks shift in the direction of the bulk plasmon peak of silicon (see Figure 3).

The removal or decrease of the low-energy loss peaks is expected to increase the secondary electron yield. By allowing the extra silicon atoms to form clusters and removing (part of) the dangling bonds, (part of) the lattice defects are removed. This decreases the probability for an electron travelling in the material to scatter at such a defect. The same holds for amorphous $Si_7N_7$. In this case, the material is allowed to relax by which (part of) the lattice defects are removed naturally. This is a more effective way of removing the lattice defects, as is seen in the energy loss function, where there are fewer energy loss peaks present for amorphous $Si_7N_7$ than for clustered $Si_7N_7$. The simulated secondary electron yield is expected to increase when increasingly more lattice defects are removed, which is seen indeed in Figure 4. Going from $Si_7N_7$ with homogeneously distributed silicon atoms to clustered silicon atoms, the maximum secondary electron yield increases from 2.67 at 350 eV to 2.72 at 350 eV, which is a small increase of 1.9%. The increase in maximum secondary electron yield for amorphous $Si_7N_7$ is more prominent; the yield increases to 2.9 at 350 eV, a total increase of 8.5%.

## 4. Conclusions

Si doping can counter the effects of charging, although a higher doping level leads to more lattice defects and a decrease in secondary electron yield. The dopant distribution can help to increase the yield slightly by the removal of lattice defects. An optimum needs to be found.

Increasing the doping level leads to a decrease in maximum secondary electron yield of 34% for $Si_{13}N_{15}$ and 46% for $Si_7N_7$ compared to $Si_3N_4$. The extra silicon atoms in SRSN introduce dangling bonds that cause the secondary electron yield to decrease. The exact distribution of these extra silicon atoms determines how much the secondary electron yield decreases. When the extra silicon atoms are placed in clusters in $Si_7N_7$ to remove some of the dangling bonds, the maximum secondary electron yield increases by 1.9% compared to homogeneous $Si_7N_7$. A more effective way to remove the lattice defects is to let the SRSN relax to amorphous $Si_7N_7$. Then the maximum secondary electron yield increases by 8.5% compared to homogeneous $Si_7N_7$. Note that this is still lower than the maximum secondary electron yield of $Si_3N_4$ and $Si_{13}N_{15}$.

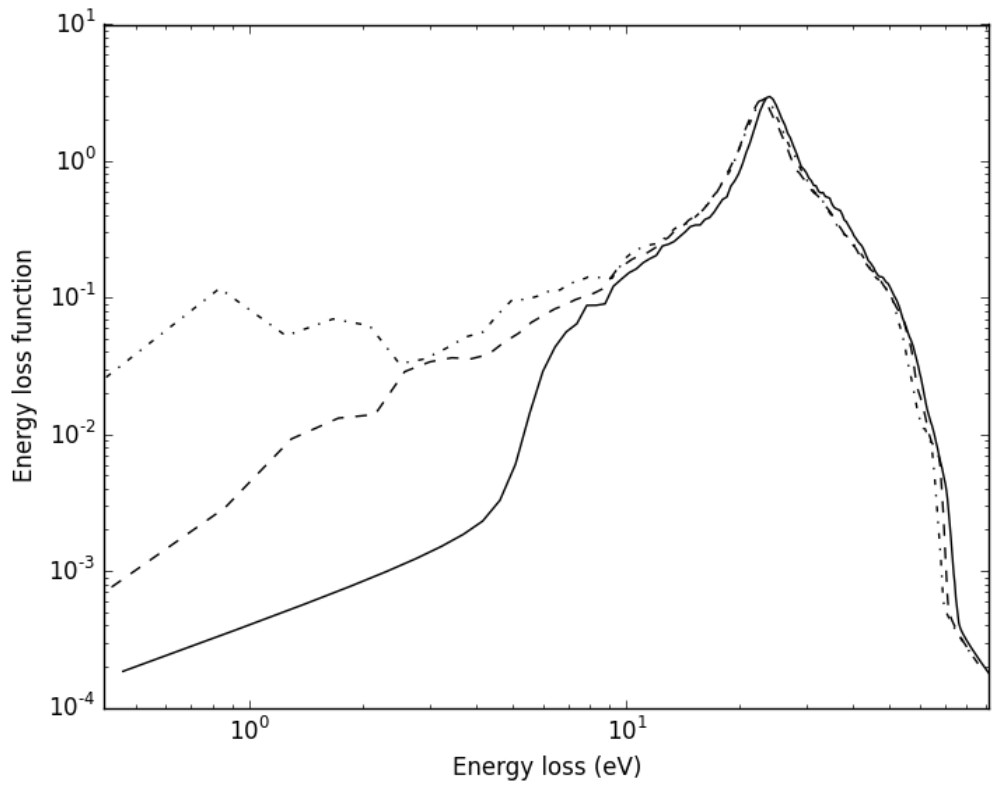

**Figure 1.** The calculated energy loss functions for $Si_3N_4$ (solid line), $Si_{13}N_{15}$ (dashed line) and $Si_7N_7$ (dash-dotted line).

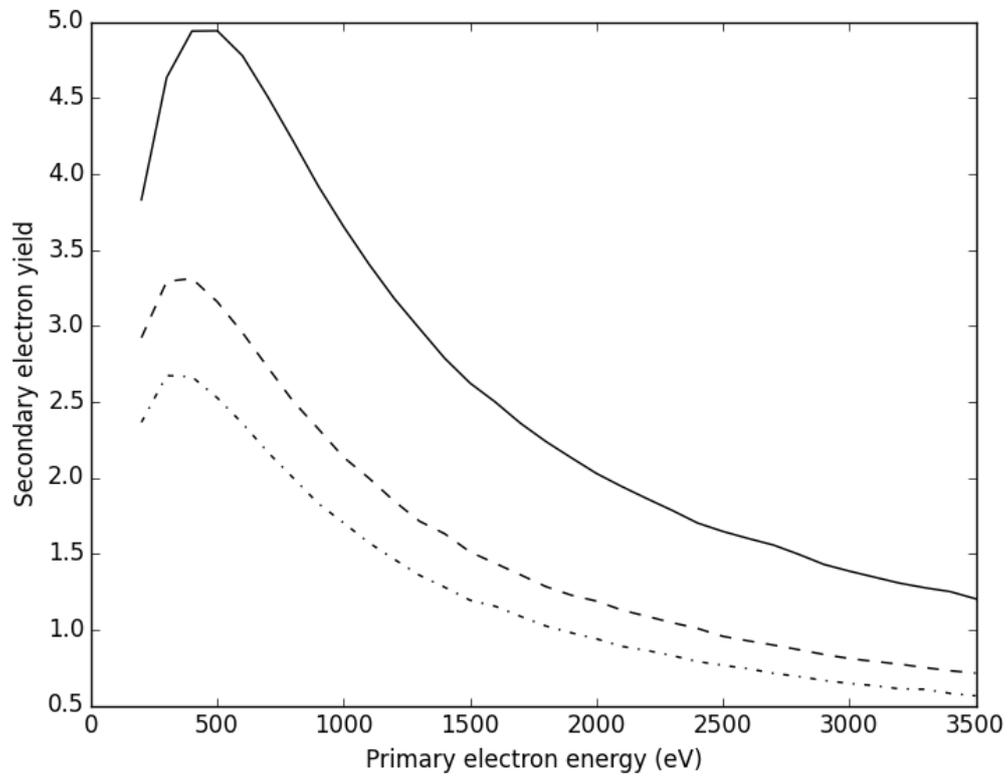

**Figure 2.** The simulated reflection secondary electron yield for $Si_3N_4$ (solid line), $Si_{13}N_{15}$ (dashed line) and $Si_7N_7$ (dash-dotted line).

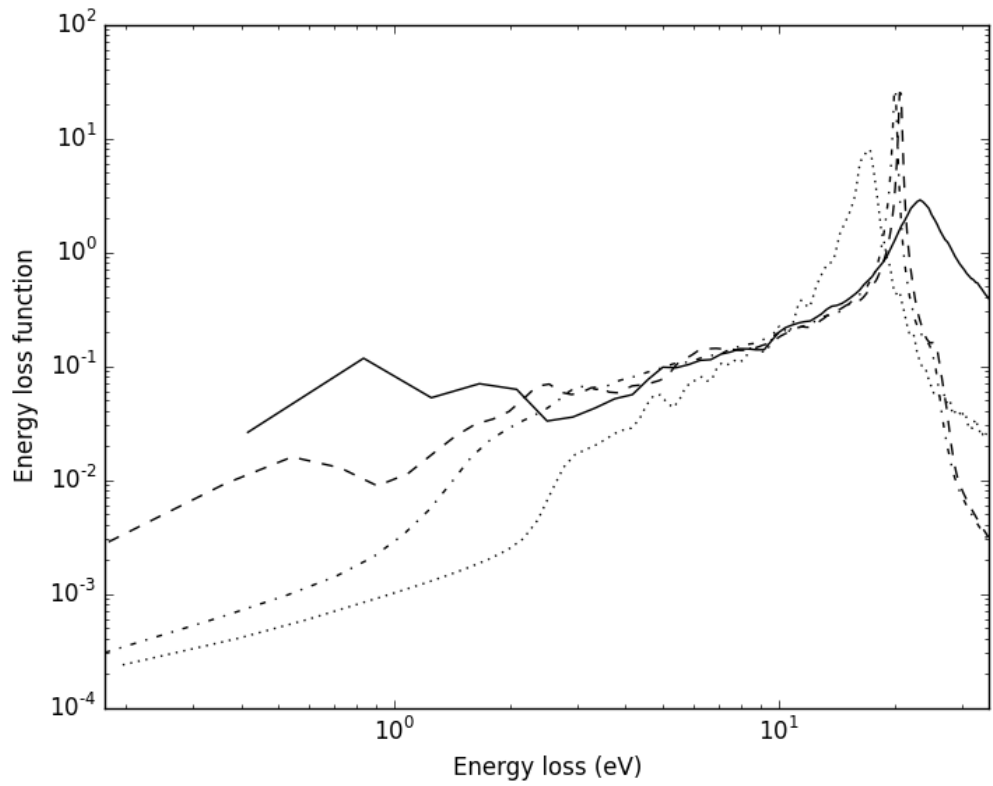

**Figure 3.** The energy loss functions for the different structures of $Si_7N_7$. The extra silicon atoms can be distributed homogeneously (solid line) or in clusters (dashed line) over the material, or the material as a whole can be amorphous (dash-dotted line). The energy loss function for silicon is added as reference (dotted line).

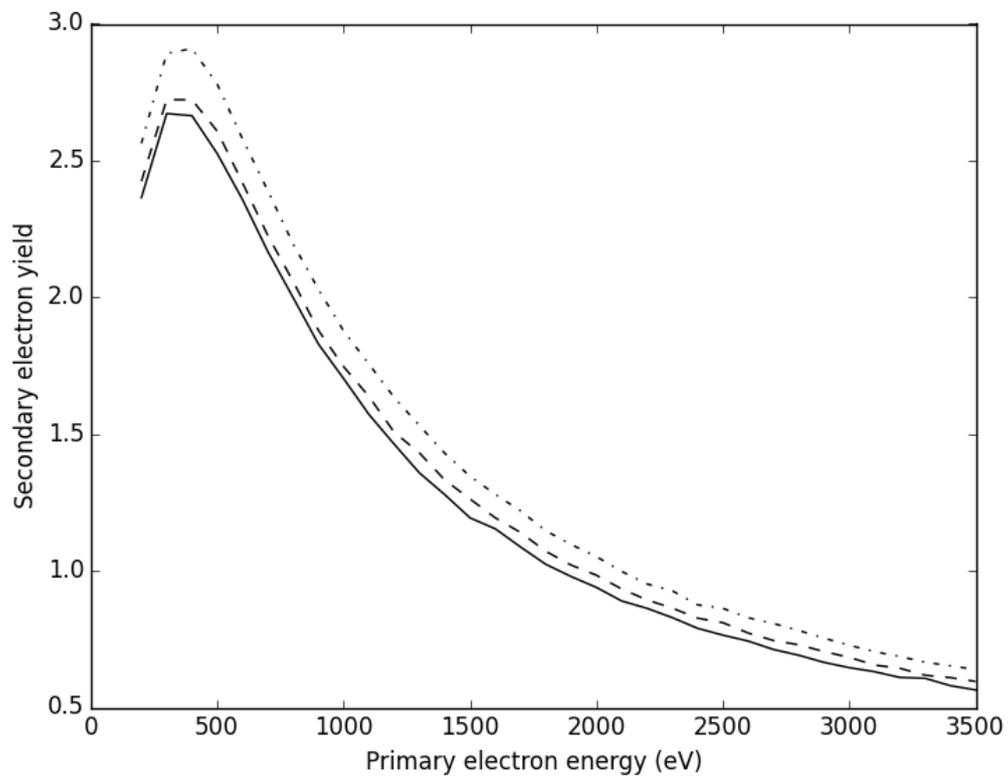

**Figure 4.** The simulated reflection secondary electron yield for $Si_7N_7$ with the extra silicon atoms distributed homogeneously (solid line) and in clusters (dashed line) over the material, and the yield for amorphous $Si_7N_7$ (dash-dotted line).